\documentclass{sigchi}

\toappear{\scriptsize Permission to make digital or hard copies of all or part of this work for personal or classroom use is granted without fee provided that copies are not made or distributed for profit or commercial advantage and that copies bear this notice and the full citation on the first page. Copyrights for components of this work owned by others than ACM must be honored. Abstracting with credit is permitted. To copy otherwise, or republish, to post on servers or to redistribute to lists, requires prior specific permission and/or a fee. Request permissions from permissions@acm.org. \\
{\emph{CHI '20, April 25--30, 2020, Honolulu, HI, USA.} } \\
\copyright~2020 Association for Computing Machinery. \\
ACM ISBN 978-1-4503-6708-0/20/04\ ...\$15.00. \\
http://dx.doi.org/10.1145/3313831.3376843 }

\usepackage{balance}       
\usepackage{graphics}      
\usepackage[T1]{fontenc}   
\usepackage{txfonts}
\usepackage{mathptmx}
\usepackage[pdflang={en-US},pdftex]{hyperref}
\usepackage{color}
\usepackage{booktabs}
\usepackage{textcomp}
\usepackage{adjustbox}
\usepackage{tabularx}
\usepackage{subcaption}

\usepackage{microtype}        
\usepackage{ccicons}          

\usepackage{todonotes}
\usepackage{times}
\usepackage{tabularx}
\usepackage{latexsym}
\usepackage{adjustbox}
\usepackage{multirow}
\usepackage{url}
\usepackage{xcolor}
\usepackage{hyperref}
\usepackage{lipsum}
\usepackage{colortbl}
\usepackage{balance}
\definecolor{mygray}{gray}{.9}

\def\plaintitle{SIGCHI Conference Proceedings Format}

\def\emptyauthor{}
\def\plainkeywords{Empirical study that tells us about people; Text/Speech/Language; Behavior Change; Crowdsourced.}

\makeatletter
\def\url@leostyle{%
  \@ifundefined{selectfont}{
    \def\UrlFont{\sf}
  }{
    \def\UrlFont{\small\bf\ttfamily}
  }}
\makeatother
\def\pprw{8.5in}
\def\pprh{11in}

\definecolor{linkColor}{RGB}{6,125,233}
\hypersetup{%
  pdftitle={\plaintitle},
  pdfauthor={\emptyauthor},
  pdfkeywords={\plainkeywords},
  pdfdisplaydoctitle=true, 
  bookmarksnumbered,
  pdfstartview={FitH},
  colorlinks,
  citecolor=black,
  filecolor=black,
  linkcolor=black,
  urlcolor=linkColor,
  breaklinks=true,
  hypertexnames=false
}

\begin{document}

\title{Effects of Persuasive Dialogues:\\ Testing Bot Identities and Inquiry Strategies}

\numberofauthors{3}
\author{%
  \alignauthor{Weiyan Shi\\
    \affaddr{University of California, Davis}\\
    \affaddr{Davis, CA, USA}\\
    \email{wyshi@ucdavis.edu}}\\
  \alignauthor{Xuewei Wang\\
    \affaddr{Carnegie Mellon University}\\
    \affaddr{Pittsburgh, PA, USA}\\
    \email{xueweiwa@andrew.cmu.edu}}\\
  \alignauthor{Yoo Jung Oh\\
    \affaddr{University of California, Davis}\\
    \affaddr{Davis, CA, USA}\\
    \email{yjeoh@ucdavis.edu}}\\
  \alignauthor{Jingwen Zhang\\
    \affaddr{University of California, Davis}\\
    \affaddr{Davis, CA, USA}\\
    \email{jwzzhang@ucdavis.edu}}\\
  \alignauthor{Saurav Sahay\\
    \affaddr{Intel Labs}\\
    \affaddr{Santa Clara, CA, USA}\\
    \email{saurav.sahay@intel.com}}\\
  \alignauthor{Zhou Yu\\
    \affaddr{University of California, Davis}\\
    \affaddr{Davis, CA, USA}\\
    \email{joyu@ucdavis.edu}}\\
}

\maketitle

\begin{abstract}


Intelligent conversational agents, or chatbots, can take on various identities and are increasingly engaging in more human-centered conversations with persuasive goals. However, little is known about how identities and inquiry strategies influence the conversation's effectiveness. We conducted an online study involving 790 participants to be persuaded by a chatbot for charity donation. We designed a two by four factorial experiment (two chatbot identities and four inquiry strategies) where participants were randomly assigned to different conditions. Findings showed that the perceived identity of the chatbot had significant effects on the persuasion outcome (i.e., donation) and interpersonal perceptions (i.e., competence, confidence, warmth, and sincerity). Further, we identified interaction effects among perceived identities and inquiry strategies. We discuss the findings for theoretical and practical implications for developing ethical and effective persuasive chatbots. Our published data, codes, and analyses serve as the first step towards building competent ethical persuasive chatbots. 

  

\end{abstract}




\section{Introduction}

As chatbots become increasingly intelligent and human-like, understanding the interaction dynamics between users and chatbots is essential. Chatbots can take on various forms of identities such as adopting a human name (e.g., Alexa, Watson) or directly disclosing itself as a bot. Regarding contents, the capacity of chatbots in conversing in natural language and performing user-targeted personalization has been greatly improved. Despite that chatbots have been rapidly deployed in business and healthcare in recent years, our knowledge regarding its potentials, impacts, and ethical standards is very limited. Recent studies on human-bot interactions suggest that bot identities and conversational strategies can significantly influence how users respond to chatbots' messages and requests. For instance, \cite{go2019humanizing} 
compared the effectiveness of a chatbot adopting two different identity cues (i.e., human or chatbot) and measured participants' perceptions toward the chatbot as well as their subsequent behavioral intention of returning to a given website. They found the most effective chatbot was the one with a human identity and delivered contingent messages. Other research demonstrated that user-targeted personalized dialogue systems can achieve better user engagement  \cite{yu2016user}. 
One approach tested by \cite{shi2018sentiment} 
suggests harnessing user sentiment information can enable the chatbot to be more user-adaptive. 

One significant factor in shaping effectiveness is users' perceived conversation quality. A high quality conversation is often derived from a positive perception towards the conversing partner, on top of fluent and meaningful message exchanges. As we move from basic assistant-like chatbots (e.g., answering user questions and helping with simple tasks) to more advanced chatbots with specific goals and inquiries (e.g., persuasive bots selling products or facilitating behavior changes), the user perception towards the chatbot or the user-chatbot relationship becomes more influential. In conversing with persuasive chatbots, people need to infer the bots' intent by interpreting the bots' inquiries and evaluate the persuasive messages' quality against their own ideas. Such complexity requires the Human Computer Interaction (HCI) community to ask more intricate questions regarding how users react to persuasive chatbots that can increasingly adopt more goal-oriented conversations. Guided by the Computers Are Social Actors (CASA) framework and the Uncanny Valley of Mind (UVM) theory, in this study, we test the effects of bot identities and inquiry strategies in a chatbot-human persuasion context. 

The primary goal of the study is twofold. First, to test the questions in a persuasive conversation context, we build our own agenda-based persuasive dialogue system utilizing the {\scshape PersuasionForGood}  dataset \cite{wang2019persuasion}. We adopt the persuasive strategy scheme and extend the human-human conversations to a human-bot setting, by developing a persuasive chatbot that uses neural network models to understand users and retrieve previously collected human responses to construct system replies.
Second, we design an online experiment that uses the chatbot to persuade human participants to donate to a children's charity. To test the effects of the chatbot identity and inquiry strategies on persuasive and interpersonal outcomes, we conduct a 2 (chatbot identity disclosure: Jessie vs. Jessie [bot]) x 4 (inquiry strategy: personal + non-personal inquiry vs.  personal inquiry vs.  non-personal inquiry vs.  no inquiry) factorial experiment and recruit a total of 790 participants. We draw on three theoretical frameworks to explain the interactions between chatbot identities and inquiry strategies on influencing relationship perceptions and persuasive outcomes. ``Computers Are Social Actors" (CASA) paradigm, ``Question-Behavior" effect, and ``Uncanny Valley of Mind" (UVM) are introduced in the next section to inform our hypothese formulation. 

Results reveal that if people perceive the chatbot as human, they are more likely to donate at the end. This main effect of perceived human identity on donation is notably salient when personal inquiries are given to participants. The main effect of human identity on increasing donation holds the same for those participants whose perceived chatbot identity aligns with the displayed identity. Importantly, participants in the human identity condition perceiving the chatbot as human show the highest likelihood of donation, whereas those in the human identity condition but perceiving the chatbot as bot show the lowest donation probability, confirming the UVM prediction. In regards to interpersonal outcomes, results reveal that when the chatbot is perceived as human, participants are more likely to rate the chatbot as competent, confident, sincere, and warm. 

Our work offers several contributions to the field of persuasive design in HCI. First, we offer design frameworks for building a persuasive chatbot with our codes, datasets, and study findings. The code and data are available \href{https://github.com/wyshi/Persuasive-Dialogues}{here}.
 Second, we stress the importance of examining users' perceptions toward the chatbot identity by showing differential effects on persuasion outcomes, illuminated by theories on human-bot interactions. Lastly, we address the ethical consideration in disclosing bot identities and intentions in light of the California's Autobot Law. 


\section{Related Work and Hypotheses}
\subsection{Effect of chatbot identity on donation}
CASA paradigm suggests that the processes that lead to user perceptions toward the conversation partner do not differ depending on the perceived identity of the partner. Regardless of whether the partner is assumed as a machine or a human, people may hold similar perceptions. This is because intelligent machines are perceived to be able to follow social norms and behave in a socially desirable manner \cite{waytz2010sees}. 
Accordingly, people apply social rules, norms, and expectations when they interact with computers \cite{lee2010trust}
. For instance, when interacting with a personified agent, people adhere to the same social conventions and rules that are used when interacting with other humans \cite{nass2000machines}
. 
CASA paradigm suggests chatbots and human partners will elicit similar perception and conversational outcomes. For instance, a study showed Twitter users' perceptions (e.g., credibility, attractiveness, communication efficiency) toward bot accounts were similar to those of human accounts \cite{edwards2014bot}
. \cite{ho2018psychological} 
studied the effects of identity (chatbot vs. human) on disclosures and found that chatbots and humans were equally effective at generating emotional, relational, and psychological benefits. Hence, we speculate that the chatbot and human identity will yield similar interpersonal and persuasion outcomes. 

Hypothesis 1: Both identities (chatbots or human) yield equivalent persuasive and interpersonal outcomes.

\subsection{Persuasive Inquiries' Effect on Donation}

Persuasive inquiries are important in eliciting persuasion outcomes. Previous research has shown that even simply asking questions about a behavior can lead to changes in the behavior, known as the ``question-behavior" effect \cite{sherman1980meaning}
.  For instance, \cite{williams2006simply} 
found that asking questions about exercise led to an increase of exercise. 
This effect is particularly salient on socially normative behaviors such as donation \cite{sprott2006question}
. For example, \cite{godin2008asking} 
demonstrated that merely asking questions regarding blood donation could promote such behavior. 

Building upon this literature, in this paper, we propose that personal inquiries would be more persuasive than non-personal inquiries, as research on personalized persuasion stresses the importance of asking personal questions in order to ``tailor" the information, arguments, or feedbacks to effectively persuade their partners \cite{lee2017human}. 
Specifically, we expect that personal inquiries (i.e., personal inquiry alone and personal plus non-personal inquiry) will yield better persuasive and interpersonal outcomes than non-personal inquiries (i.e., non-personal and no inquiry). Thus, we propose:

Hypothesis 2: Personal inquires will yield greater persuasive and interpersonal outcomes than non-personal inquiries.

\subsection{Interaction between Identity and Inquiries on Donation}

Until now, little research has investigated the interaction effect of chatbot identity and persuasive inquiries on persuasion outcomes. In this study, we adopt the UVM theory to inform our thinking. Unlike the CASA framework, UVM suggests that perceptual difficulty in discerning a human-like object and its human characteristics will evoke negative (e.g., eerie, uncomfortable) feelings \cite{mori1970electron}
. That is, excessive similarities may elicit repulsive and negative reactions from the part of the user \cite{de2001unfriendly}
. As robots reach near-human like forms, the uncanny valley effect becomes salient. For instance, a study  showed that participants experienced more negative feelings with a complex animated avatar chatbot than with a simpler text-based chatbot \cite{przegalinska2018muse}
. 
The original UVM did not predict the eerie feeling triggered by a human agent (perceived as human) exhibiting robot characteristics (e.g., robotic appearances or demeanor). However, recent studies argue that the opposite may elicit eerie feelings as well
. \cite{gray2012feeling} claimed that the effect can apply beyond machines, such that a human lacking experience (e.g., psychopaths) could also elicit the uncanny valley effect. For example, in \cite{corti2015revisiting}
, they asked the participants to speak with a partner who was assumed to be a human but actually speech shadowing for an AI chatbot. Results showed that the participants expressed discomfort towards their partners. In sum, the current UVM literature argues that the uncanny valley may apply to both cases: human-like bots and robotic humans \cite{gray2012feeling}
. Therefore, we expect that conversing with a chatbot that is perceived as a human would elicit perceptions of ``uncanniness''. We speculate that if the chatbot tries to interact in a personal and human-like way (e.g., by asking personal questions), people may feel uncomfortable, which can subsequently degrade the interpersonal perceptions of the partner as well as their persuasiveness, as hypothesized below: 

Hypothesis 3: There is an interaction effect between chatbot identity and persuasive inquiry type on persuasive and interpersonal outcomes.

\subsection{Persuasive Strategies}
Persuasive systems are special dialogue systems that attempt to change people's mind or behavior by employing various {\it persuasive strategies} \cite{fogg2002persuasive, Gold1992, sale, tutor, oinas2008towards}. Various persuasive strategies have been well documented \cite{hidey2018persuasive, hidey2017analyzing, yang2019let, hibbert2007guilt, tan2016winning, oinas2009persuasive}. For example, \cite{hidey2017analyzing} proposed a two-tiered annotation scheme to distinguish claims and premises and their semantic types in an online persuasive forum; \cite{hibbert2007guilt} analyzed the guilt appeal in a charitable giving setting and \cite{oinas2009persuasive} presented 28 persuasive strategies. For this study, we choose to adopt the persuasive strategy scheme and
the {\scshape{PersuasionForGood}} dataset from \cite{wang2019persuasion} to develop the dialogue system to deploy a similar task, and to extend the human-human conversations in a human-bot setting. 
In {\scshape{PersuasionForGood}}, there are 10 persuasive strategies that can be categorized into two groups, seven \textit{persuasive appeals} and three \textit{persuasive inquiries}. Persuasive appeals try to change donation decisions whereas persuasive inquiries attempt to facilitate personalized contents and closer relationships by asking questions.
Here are the seven persuasive appeals' definition from \cite{wang2019persuasion}, along with example sentences listed in  the supplementary material: 1) {\it credibility appeal} refers to the uses of organization facts to establish credibility and gain trust, e.g. ``The organization is highly rated with many rewards''; 
2) \textit{emotional appeal} refers to the elicitation of certain emotions, such as empathy or guilt to influence people, e.g. ``Kids are suffering from hunger''; 3) \textit{ logical appeal} refers to the uses of reasoning and evidence, e.g. ``Your donation could go to address this problem and help many children''; 
4) \textit{self-modeling} convinces people by  indicating the persuader's positive donation intention and acting as a role model, e.g. ``I will donate \$2 myself''; 5)  \textit{foot-in-the-door} refers to the strategy that starts with small donation requests to encourage compliance followed by larger requests \cite{foot}, e.g. ``even \$0.01 would help a lot''; 6) \textit{personal story} uses narrative exemplars to demonstrate someone's donation experiences to motivate others to follow, e.g. ``My brother and I replaced birthday gifts a few years ago''; 7) \textit{donation information} provides donation task details to increase the persuadee's self-efficacy to complete donations, e.g. ``Your donation will be directly deducted from your task payment''.







Following \cite{wang2019persuasion}, we separate persuasive inquiries into non-personal and personal ones. Table~\ref{tab:sample sentence for inquiry} shows some examples.

\begin{itemize}
    \item \textbf{Non-personal Inquiry} refers to relevant questions without asking personal information. It include two sub-categories: 1)  source-related inquiry that asks if the persuadee is aware of the organization, and 2) task-related inquiry that asks the persuadee's opinion and experience related to the donation task.   
    \item \textbf{Personal Inquiry} asks about persuadee's personal information relevant to donation for charity but not directly on the task, such as ``Do you have kids?''
\end{itemize}


\begin{table}[h]
\small
\begin{adjustbox}{width=\columnwidth}
\begin{tabular}{l|l}
\hline

\textbf{Persuasive Inquiry}   & \textbf{Example} \\\hline

Non-personal inquiry & \textit{\begin{tabular}[c]{@{}p{55mm}@{}}Have you heard of Save the Children? \\ Have you donated to a charity before? \end{tabular}} \\\hline  


Personal inquiry & \textit{\begin{tabular}[c]{@{}p{50mm}@{}}Are you aware of the dangerous situations children face in conflicted areas? \\ Do you have kids? \end{tabular}} \\\hline  

\end{tabular}
\end{adjustbox}

\caption{ Templates for the two types of persuasive inquiries 
}
\label{tab:sample sentence for inquiry}
\end{table}

\section{Persuasive Dialogue System}
In \cite{wang2019persuasion}, the donation task involved two humans, one persuader and one persuadee. This dataset is suitable to model human persuasion and to test the difference between human-bot and human-human conversations.
Utilizing the {\scshape{PersuasionForGood}} dataset, we built an agenda-based persuasive dialogue system to interact with human users. Agenda-based dialogue system follows certain pre-defined rules and is easy to control in experimental conditions. The system consists of three components: 1) the Natural Language Understanding (NLU) module that interprets the user sentences into a condensed dialogue-act (e.g. ask-org-info), 2) the Dialogue Manager (DM) module that takes in user dialogue-act and decides the system's next dialogue-act (e.g. provide-org-info); 3) the Natural Language Generation (NLG) module that converts the previously-determined system dialogue-act to human-readable sentences, as responses presented to the user. Each module was trained with the  {\scshape{persuasionforgood}} dataset. We now describe the modules in detail.
 
\begin{table}[h]
\small
\begin{adjustbox}{width=\columnwidth}

\centering
\begin{tabular}{lp{71mm}}
\bf Role & \bf Utterance\\\rowcolor{mygray}

Jessie   &  Hello! How are you doing today?  \\\rowcolor{white}
User   & Hi there! Doing well and how about yourself?\\\rowcolor{mygray}
Jessie   &  I am doing good, thanks for asking. Have you heard of the organization "Save the Children"?  \\\rowcolor{white}
User   & I have heard a bit about them. I would love to know more. Do you know much about the organization?\\\rowcolor{mygray}
Jessie   &  Save The Children is an international non-governmental organization that promotes children's rights, provides relief and helps support children in developing countries. Have you donated to a charity before?  \\\rowcolor{white}
User   &I donate to my local rescue mission as well as my local animal shelter. \\\rowcolor{mygray}
Jessie   &  Are you aware of the dangerous situations children face in conflicted areas?  \\\rowcolor{white}
User   & I am somewhat aware and it greatly bothers me\\\rowcolor{mygray}
Jessie   & Do you have kids yourself?   \\\rowcolor{white}
User   & Yes I am a parent of two children and I hate to think of any child suffering.\\\rowcolor{mygray}
Jessie   & Do you want to make a donation to "Save the Children"? I think the refugee crisis is crazy, I am not sure what I would do if I was raising my kids in Syria. If anyone could have fun in those spots it's kids though. You can choose any amount from \$0 to all your task payment (\$2) to donate, which will be directly deducted from your payment. After task completion, the research team will send the donations to "Save the Children".   \\\rowcolor{white}
User   & Yes I would like to donate \$1\\\rowcolor{mygray}
Jessie   &  Thanks so much! Any chance you would consider making a greater donation?  \\\rowcolor{white}
User   & I would like to make it \$2 actually!\\\rowcolor{mygray}
Jessie   &  Thank you so much! Just to confirm, how much would you like to donate?  \\\rowcolor{white}
User   & I would like to donate \$2 to "Save the Children"\\\rowcolor{mygray}
Jessie   &  Thank you. It's been lovely talking to you. Enjoy your day and bye!  \\\rowcolor{white}
User   & Thankyou and it was great chatting with you as well! \\\rowcolor{mygray}
Jessie   & I think you can now click the "Done" button below to the next page.    \\\rowcolor{white}
User   & I am done with the chat and clicked the 'Done' button, thank you! 
\end{tabular}

\end{adjustbox}
\caption{\label{tb:0}One example  persuasion dialogue. ``Jessie'' and ``User'' refer to the displayed bot identity and  the human user. The user donated \$2 after the conversation.}
\end{table}

\subsection{Natural Language Understanding (NLU)}
The input to the NLU module is the user utterance, and the output is a user dialogue-act from a pre-defined set of dialogue-acts. This set is a subset of user dialogue-acts from \cite{wang2019persuasion} including 14 essential dialogue-acts, such as \textit{agree-donation} and \textit{ask-org-info}. 
We adopted the persuasion strategy classifier in \cite{wang2019persuasion} and trained it with the user utterances in {\scshape PersuasionForGood} to predict the user dialogue-act. Specifically, we combined 4 types of features to perform the classification task, 1) the user utterance itself, 2) the utterance textual feature such as sentiment scores, 3) the contextual feature from the previous utterance and 4) the utterance character feature.
First, we passed the persuadee's utterance word embeddings \cite{Fast} into two-layer RCNN \cite{Lai2015}  to obtain the utterance representation. 
We utilized hand-crafted textual, contextual and character features to improve the model's performance. Textual feature consists of sentence sentiment scores obtained from the open-source toolkit VADER \cite{vader}; the sentiment score would help the classification because certain utterances that belong to certain dialogue-act such as ``reaction-to-donation''  contain emotional words. Context information was the output from the CNN model with the previous persuader's utterance as input, which can be viewed as a context representation. Character information was calculated by a pretrained char embedding \cite{review}. Finally,
we concatenated the utterance embedding along with these three additional features to train the RCNN user dialogue-act classifier. We trained the model on the training dataset and fine-tuned it on the validation dataset. Our model reached 62\% in  accuracy on the test set. The accuracy is not high, because we have 14 classes and highly-unbalanced dataset. To improve the NLU performance, we also patched the automated  classifier with regular expressions and pre-defined rules to reach 84.1\% in accuracy. 

\begin{figure}[htb!]
\centering
  \includegraphics[width=0.9\columnwidth]{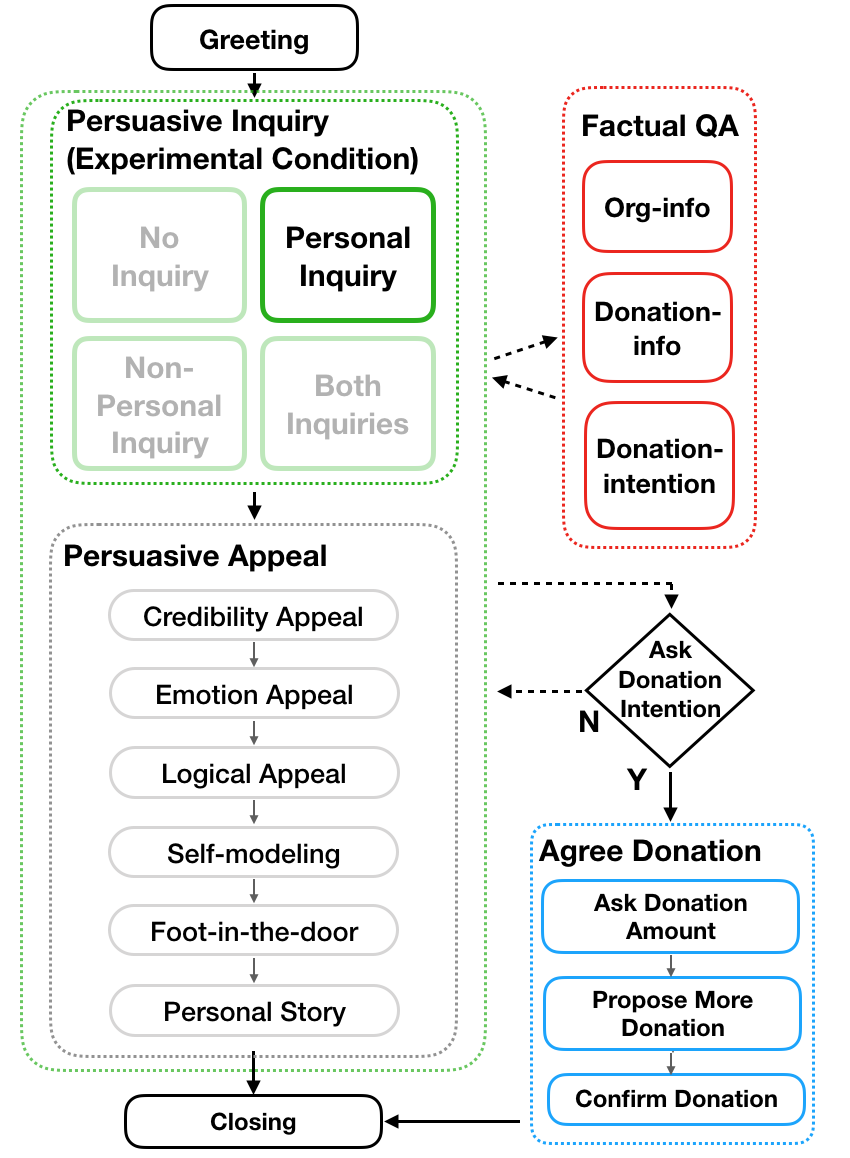}
  \caption{The system agenda. }~\label{fig:agenda}
\end{figure}

\subsection{Dialogue Manager (DM)}
The Dialogue Manager outputs the system dialogue-act given a user dialogue-act. It follows the agenda shown in Figure~\ref{fig:agenda}. The system first greets the user, and then enters the ``Persuasive Inquiry'' module in green and starts to persuade. According to the experimental condition, the system chooses different type of persuasive inquiries to ask the user different information. For example, if the user is assigned to the ``personal inquiry'' condition, the system will ask personal information. After the ``Persuasive Inquiry'' module, the system proceeds to the ``Persuasive Appeal'' module in grey where different appeal strategies are used in a fixed order. The reason we fixed the appeal order is that previous study \cite{wang2019persuasion}  didn't find a  significant interaction effect between persuasive appeals and personalization. Therefore, the order of persuasive appeals should be a controlled variable. The appeal order was chosen by the recommended optimal order in \cite{wang2019persuasion}. Specifically, {\it credibility appeal} appears in the beginning, followed by  {\it emotional appeal} and {\it logical appeal}, then {\it self modeling} is presented, and the last two are {\it foot-in-the-door} and {\it personal story}. {\it Donation information} always follows the first propose-donation dialogue act, and appears whenever the user asks about the donation procedure. 

At every turn, the NLU module predicts the user dialogue-act. Whenever the predicted user dialogue-act is ``agree-donation'', the system enters ``Agree Donation'' module (in blue) with a sequence of three system dialogue-acts: 1) The system will ask the exact donation amount first if it hasn't been provided; 2) Then the system will propose a bigger donation by asking ``Any chance you would consider making a greater donation?' 3) Next, the system will confirm the donation amount with the user and then close the conversation. If the user doesn't agree to donate throughout the conversation, after the system runs out of persuasive appeals, it will close the conversation. If the user has already finished 10 turns of conversations, the user has the option to continue or exit the dialogue.

In addition, there is a ``Factual QA'' module (in red) to answer user-initiated questions on  ``organization-information'', ``donation-information'' and ``self-donation-intention (users ask the persuader if they will donate)''. For example, whenever the user asks about the organization facts during the conversation, the system will jump out from the previous module and enter the ``org-info'' in the ``Factual QA'' module to provide answers to the user's question. After that, the system will go back to the previous module and retrieve an appropriate response to lead the conversation. Then, the two utterances are concatenated to be the system response. The same procedure is also applied to the ``donation-info'' and ``self-donation-intention''.

\subsection{Natural Language Generation (NLG)}
After the dialogue manager chooses the system dialogue-act (e.g.{\it propose-donation}), we need to generate human-readable language from the previously chosen dialogue-act. There are three main approaches to perform NLG: the template-based method that has a fixed set of human-written templates for each dialogue-act, the retrieval-based method that selects candidates from an existing corpus, and the generation-based method that  directly generates the response with a language model trained from an existing corpus.

Persuasive messages are human-readable surface-forms of the persuasive strategies, and messages from the same persuasive strategy take various surface-forms, which may have impact on the persuasive outcomes. In this paper, we want to study the effects of different persuasive inquiries instead of the impact of the surface-form; therefore, the surface-forms of the persuasive inquiries should be a controlled variable that stays the same across experiments. So we chose template-based method for the persuasive inquiries. Table~\ref{tab:sample sentence for inquiry} shows the template for each persuasive inquiry. 

For persuasive appeals, because they happen in the later part of the conversation after the persuasive inquiries, there are usually a lot more context to track and therefore, responses from template-based method would be too rigid and bland for various scenarios. In order to find more coherence responses and encourage language diversity,
we chose retrieval-based method. Since all the experiments shared the same retrieval-based NLG module for persuasive appeals and we had a closed set of existing candidate responses and large samples, there won't be a major difference between groups in terms of message quality. In this way, we also controlled the persuasive appeals' surface-form effects. Moreover, we chose to retrieve candidate responses for persuasive appeals from {\scshape Persuasionforgood} instead of  generation-based methods because 1) {\scshape Persuasionforgood} contains rich and high-quality persuasive messages, and is sufficient for retrieval; 2) The data size of {\scshape Persuasionforgood} is small relative to its rich information, which limits the generation capacity. Following \cite{he2018decoupling,shi2019build}, we calculated  the TF-IDF weighted bag-of-words vector of the last user utterance to represent the dialogue context $c_t$, and then computed the similarity score between the candidates' context vectors with $c_t$ to retrieve the system response. To avoid repetitive responses and encourage diversity, we picked the top three candidates with the highest similarity scores and randomly chose one among them.


\section{The Persuasion TASK}
Following \cite{wang2019persuasion}, we designed a similar online persuasion task about charity donation. The only difference is that in our task, the user conversed with a chatbot instead of a human to make donations to the charity ``Save the Children''\footnote{https://www.savethechildren.org/}. 
The task consisted of four parts, a pre-task survey, a dialogue, a post-task survey, and a donation confirmation. Before entering the conversation, participants were asked to complete a pre-task survey to assess their psychological profile variables, including the Big-Five personality traits \cite{Gold1992} (25 questions), the Moral Foundations endorsement \cite{Graham} (23 questions), the Schwartz Portrait Value (10 questions) \cite{Cie}, and the Decision-Making style (4 questions) \cite{Hami}. These questions are all simple multiple-choice questions with established scales. From the pre-task survey, we obtained a 23-dimension psychological profile vector where each element is the score of one characteristic, such as extrovert and agreeable. Next, the users started to talk with the persuasive chatbot. Participants were given simple instructions that they would talk about a charity, and they were informed that their donation would be directly deducted from their task payment (\$2). Such a direct deduction design made the task much more realistic than a role-play one, and users were likely to be motivated knowing their donations were real. 

Participants were randomly assigned to one of the eight experimental conditions, described in more details in the Experiment section. Participants could choose to exit the conversation after 10 conversational turns or if the system promptly closed the conversation. In the post-survey, we measured the following: 
1) attitude change towards donation, such as ``Are you going to donate to the organization in the future, if not today''; 2) the bot's quality such as persuasiveness, sincerity, and warmth; 3) bot identity, such as ``do you think you are talking to a chatbot or a human''; 4) demographic information.
We also added a simple attention-checking question ``What's the organization mentioned in the conversation''  to filter out careless users. At last, we confirmed donation by letting users privately indicate their final donation amount. Note that this amount could be different from what they promised in the conversation.

We recruited 790 participants from Amazon Mechanical Turk (AMT) and utilized ParlAI \cite{miller2017parlai} to perform the data collection. AMT has been frequently used in social science studies 
\cite{orji2018personalizing, bentley2017comparing, hirsh2012personalized, orji2017towards, orji2014modeling, buhrmester2011amazon, mason2012conducting} and is suitable for our experiments.  
To avoid repeated participants, each AMT user was only allowed to do the task once. Participants were paid USD \$2.00 each after finishing the entire task, which takes approximately 15 minutes (8 minutes for surveys and 7 minutes for conversation). All study procedures were approved by the authors' institutions. 

\section{Experiment Design and Methods}
The primary goal of this study is to answer the following questions: 1) whether there is an effect of the chatbot identity disclosure on the user's donation behavior, and 2) whether the donation behavior will differ among the various types of persuasive inquiries. We designed a 2 (chatbot identity disclosure: Jessie vs. Jessie (bot)) x 4 (persuasive inquiry strategy: personal + non-personal inquiry vs. personal inquiry vs. non-personal inquiry vs. no inquiries) factorial experiment. 
For identity disclosure, we used two identities: 1) human name ``Jessie''; 2) human name disclosed as bot ``Jessie (bot)'', to clearly inform the users that their partner was a chatbot. We chose ``Jessie'' to increase the ecological validity of the identity manipulation, following the current practice in labeling chatbot with a human-like name (e.g., Amazon's Alexa and IBM's Watson).  We chose a new name ``Jessie'' to avoid eliciting existing biases or attachments with existing names. Also, ``Jessie'' is unisex and would minimize the gender bias.
The identity labels were always shown at the beginning of every system utterance, and thus were presented at least ten times during the conversation. 
 Then, we created four inquiry conditions, where the chatbot starts the conversation in four different ways: 1) no inquiry and proposing a donation directly, 2) non-personal inquiry only, 3) personal inquiry only, and 4) both inquiries. There are two template questions for each inquiry,  listed in Table~\ref{tab:sample sentence for inquiry}; if the condition contained certain inquiries, both template questions would be used in the order listed in Table~\ref{tab:sample sentence for inquiry}. 

\subsection{Participant Demographics}
We obtained 790 unique participants, approximately 100 for each of the eight conditions after deleting invalid dialogues from non-cooperative or careless participants (e.g. typed less than 50 words during the whole conversation). Their demographic information is shown in Table~\ref{tab:demo info}. Participants came from diverse backgrounds, in terms of gender, age, education level, ethnicity, and bot familiarity. We checked the distributions of demographics and the chi-square tests showed that there was no significant difference on demographics across conditions, confirming the success of randomization.

\begin{table}[]
\small
\begin{tabular}{l|p{63mm}}
\hline
\multicolumn{2}{c}{\textbf{Total participants = 790}}                                                                          \\ \hline
\multirow{1}{*}{Gender}    & Females (60.25\%), Males (39.37\%), Others (0.38\%) \\ \hline
\multirow{2}{*}{Age}       & 18-25 (18.48\%), 26-35 (36.84\%),\\
& 36-45 (21.52\%), 45+ (23.16\%) \\ \hline
\multirow{4}{*}{Education} & Less than high school (0.51\%), High school  or equivalent (9.87\%), Some college (36.71\%), Bachelor's degree (34.56\%), Some postgraduate (4.18\%), Postgraduate agree (13.72\%) \\ \hline
\multirow{4}{*}{Ethnicity} & Asian/Asian American (6.58\%), Black/African American (9.37\%), White/Caucasian (76.20\%), Hispanic/Latino (4.05\%), Indian Subcontinent (0.51\%), Native American/American Indian (0.63\%), Other (1.78\%) \\ \hline
\multirow{2}{*}{Bot Usage} & Heavy usage (2.5\%), Moderate usage(16.6\%), Light usage (80.9\%)\\
\hline
\end{tabular}
  \caption{Participants' demographic information}~\label{tab:demo info}
\end{table}


\subsection{Dependent Variables}
Dependent variables were measured after the conversation. The complete post survey can be found in the supplementary material.
\begin{itemize}
    \item {\bf Persuasion Outcome} is measured by the donation probability. {\bf Donation Probability} is a continuous variable between 0 and 1 estimated by the relative frequency of donation within a group. The donation amount was dichotomized to indicate whether a donation was made or not (1 = donation, 0 = no donation). This is a direct metric to measure the persuasion outcome, and also a more important indicator of persuasion success than the donation amount whose variation is limited by the maximum amount of \$2. 

    \item {\bf Partner Impression} contains four dimensions: competence, confidence, warmth, and sincerity. Each is an ordinal variable with five levels, e.g. from ``incompetent'' to ``competent''.

    \item {\bf Conversation Quality} includes user engagement, response naturalness, and persuasiveness. 
{\bf Engagement} is an ordinal variable with five levels (from ``Actively disengaged'' to ``Highly engaged'') measuring how engaged the participants were during the conversation. {\bf Response Naturalness} and {\bf Persuasiveness} are both ordinal variables with five levels indicating their agreement levels to the statements ``My partner's responses are natural'', and ``I think my partner is persuasive'' respectively.
\end{itemize}

\subsection{Measurement Validation}
We performed the Kaiser-Meyer-Olkin (KMO) sampling adequacies test and the Barlett Test of Sphericity, standard tests adopted by many HCI studies \cite{orji2017towards, orji2018personalizing}, to check the data validity. The KMO was 0.78, above the recommended value of 0.6. The Bartlett Test of Sphericity showed statistical significance ($\chi ^2(561) = 9,639.27)$, $p<0.001$), ensured the data were sufficient for future analysis.

\section{Results}
We present the results in the following order:  1) persuasion outcome; 2) partner impression and conversation quality; 3) LIWC linguistics features; 4) inconsistent donation behavior; and 5) donation behavior and personality. Bonferroni correction was applied on all the p-values in the multiple t-test results shown below.

\subsection{Persuasion Outcome - Donation Behavior}
\label{sec: donation outcome}

\begin{figure*}[htb!]
\centering
\begin{minipage}[b]{.48\textwidth}
  \centering
  \includegraphics[width=0.9\columnwidth]{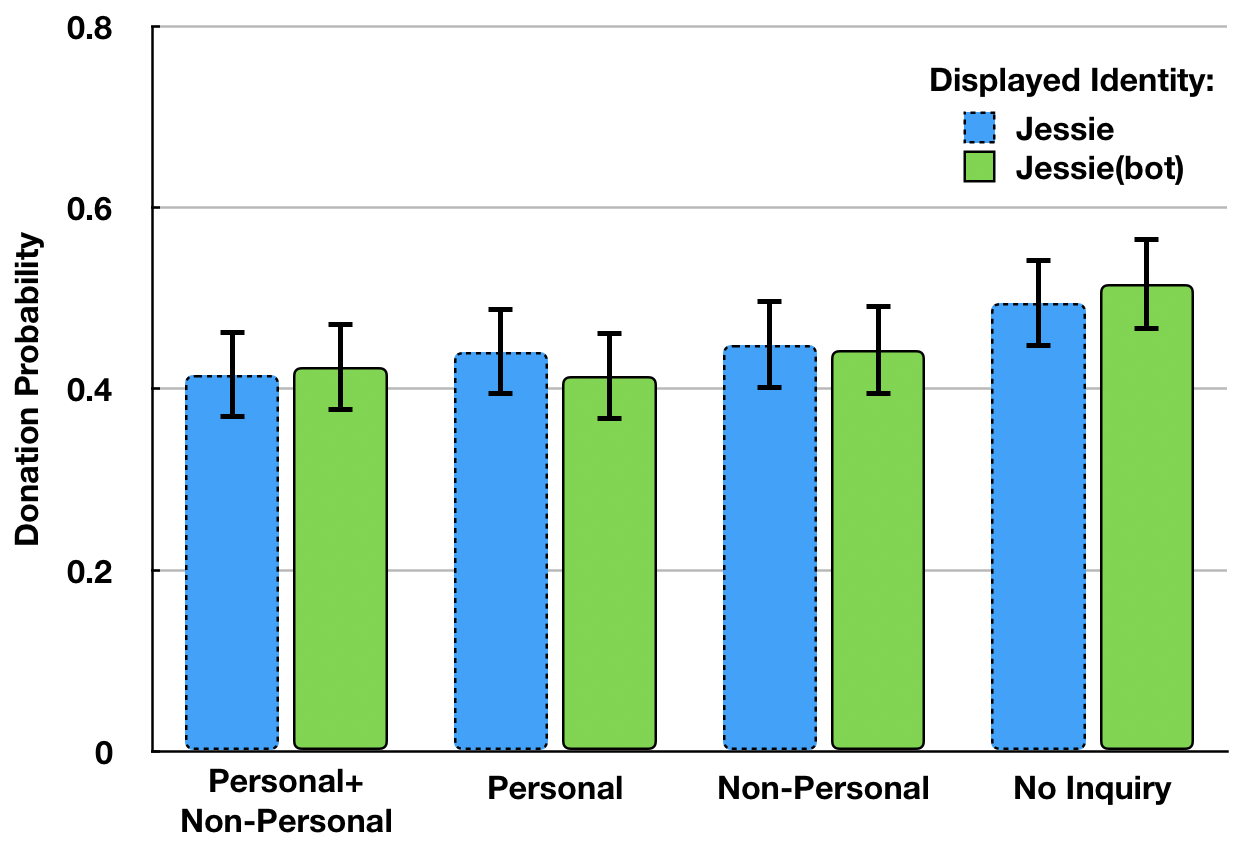}
\caption{Donation probability under two \textit{displayed identities} and four persuasive inquiries \textbf{on the whole dataset}.}
\label{fig:bot inq donation}
\end{minipage}\qquad
\begin{minipage}[b]{.48\textwidth}
  \centering
  \includegraphics[width=0.9\columnwidth]{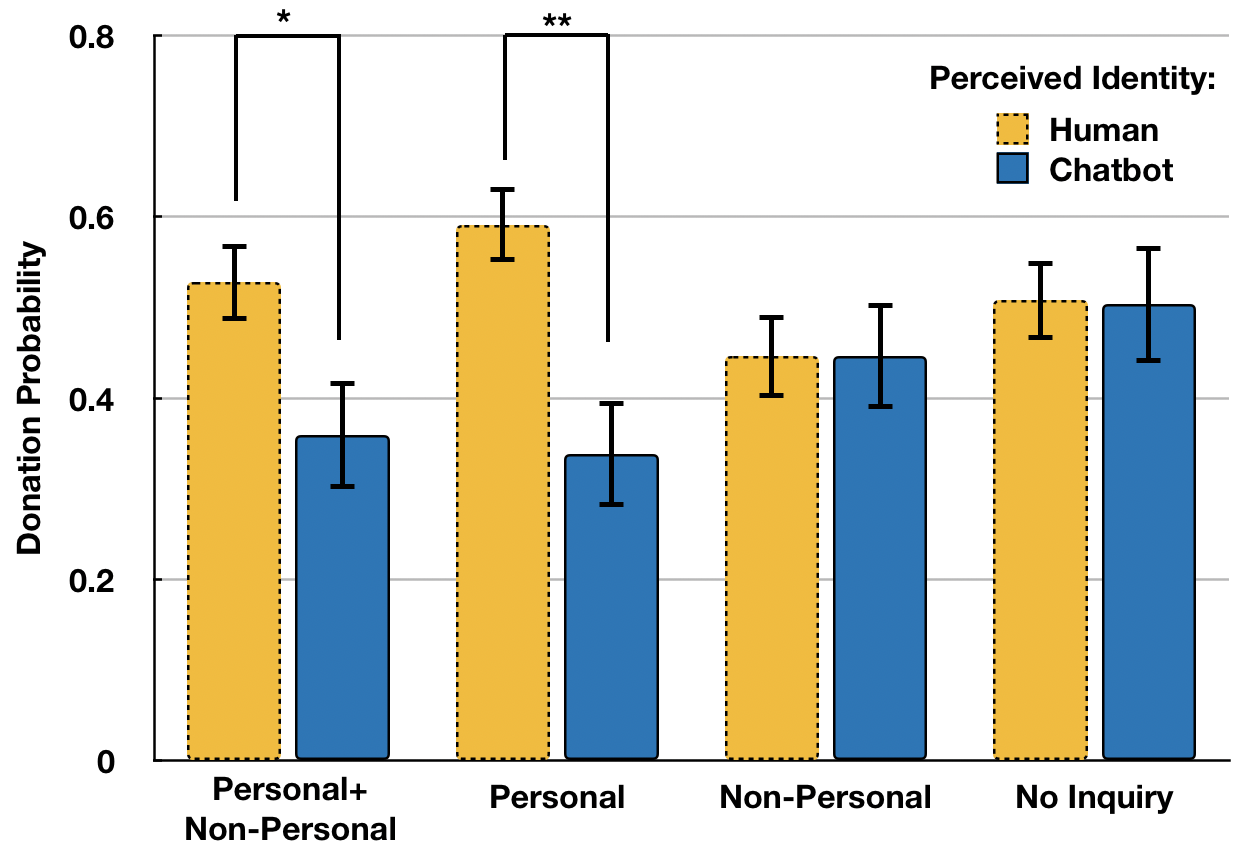}

\caption{Donation probability under two \textit{perceived identities} and four persuasive inquiries \textbf{on the whole dataset}.}
\label{fig:figure1}
\end{minipage}
\end{figure*}

Our primary dependent variable is the persuasion outcome, namely, the estimated donation probability. On average, the donation probability across all conditions was 45\%, slightly smaller than the probability of 54\% observed in the earlier research from human-human persuasions \cite{wang2019persuasion}. Figure~\ref{fig:bot inq donation} shows the donation probabilities from all experimental conditions. Although the conditions with no inquiries showed greater donations, the difference was not significant.

Considering the theoretical arguments from CASA and UVM regarding perceptions toward the conversing partners, we looked into participants' perceptions of the chatbot. In the condition where we disclosed the bot's identity as ``Jessie (bot)'', surprisingly, there were still 36.1\% participants reported that they thought their partner was a human. In contrast, 34.3\% of participants in the condition with ``Jessie" thought they were talking to a human. This suggests regardless of the label, participants had their own judgments of the bot identity. Figure~\ref{fig:figure1} showed the donation probability by perceived identity. We used the perception variable to model the donation behavior. Results from the logistic regression on donation showed a main effect of the perceived identity, such that if participants perceived the partner as a human, the donation probability increased ($\beta=0.44, p<0.01$), which contradicts the CASA paradigm in Hypothesis 1 that anticipates both chatbot and human identity to yield equivalent persuasion outcomes. Further, we identified a significant interaction between personal inquiry and perceived identity ($\beta=1.02, p<0.05$). We found the positive effect of personal inquiries in comparison to no inquiry was stronger when the participants perceived the partner as a human. This proves our Hypothesis 2 that personal inquiries will produce greater persuasion outcomes, when the partner is perceived as a human. 
Further t-tests showed that in the conditions where personal inquiries were used, and if the participants thought they were talking to a human, the donation probability would be significantly higher than if they thought they were talking to a bot ($p < 0.01$); we observed the same effect in the condition where ``both inquiries'' were applied  ($p<0.05$).  This suggests perception plays a significant role in determining inquiry effects. If participants thought they were talking to a human partner, they were more likely to be persuaded by personal inquiries. However, if they thought they were talking to a bot, personal inquiries were not effective.



\begin{figure}
\centering
  \includegraphics[width=0.9\columnwidth]{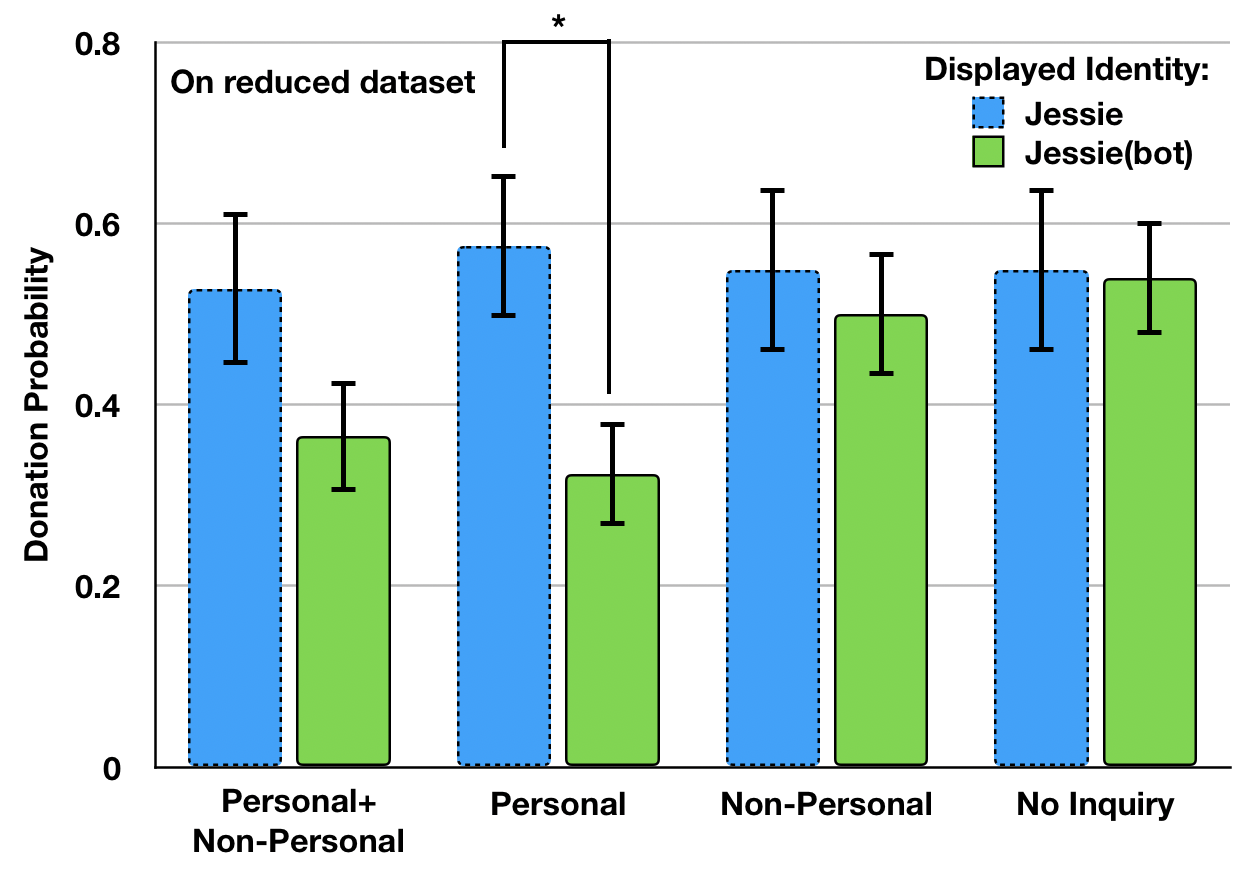}
  \caption{Donation probability under two \textit{displayed identities} and four persuasive inquiries \textbf{on the reduced dataset} which only contains users who believed in the displayed identity. }~\label{fig:reduced}
\end{figure}

\begin{table*}
\centering
\begin{adjustbox}{width=0.9\textwidth}

\begin{tabular}{|l|l|}
\hline
\textbf{I think my partner is human because...}                                                                                                                   & \textbf{Condition}    \\ \hline
My partner asked me if I was a parent myself and seemed very warm and interested in my life.                   & Jessie       \\ \hline
They reminded me to exit out of the chat.                                                                                                                     & Jessie       \\ \hline
Responded to my dialog with heartfelt responses.                                                                                                              & Jessie (bot) \\ \hline
The responses seemed natural and like other conversations I've had in the past.                                  & Jessie (bot) \\ \hline
\textbf{I think my partner is a bot because...}                                                                                                                     &              \\ \hline
I'm divided... the replies were so quick so I think chatbot, but they did have natural sounding replies.    & Jessie       \\ \hline
I think my partner was a chatbot because they wouldn't take my answer and kept asking repeatedly.                                                                                                                     & Jessie       \\ \hline
The way they phrased things made it seem like a chat bot.     & Jessie (bot)       \\ \hline
The platform clearly said they were a bot and just spouted out answers that really didn't make sense with what I was telling them.     & Jessie (bot)      \\ \hline
\end{tabular}
\end{adjustbox}
\caption{The collected reasons why the participants perceived their partners as human or bot, with the experiment condition. }
~\label{tab:reason}
\end{table*}

We also asked why participants thought their partners were human or bot in the post survey. Selected user responses are shown in Table~\ref{tab:reason}. Some participants didn't believe in the human identity because their partner ``wouldn't take the answer and kept asking repeatedly'', which could be due to  user dialogue-act classification errors. Some participants felt ``divided'' and suspected the identity displayed to them; they couldn't distinguish the identity because ``the conversation seems human like yet the replies were too quick''. Some participants believed they were talking to a human even though they were shown talking to ``Jessie (bot)'', because they thought the bot responses ``were appropriate and heartfelt''. In sum, we did observe that participants had different perceptions and suspicions of the bot's displayed identity.

These observations led us to think there might be an interaction effect between people's suspicion of the bot identity and the experimental conditions. So we defined suspicion to be a binary variable (1 = the perceived identity is different from the displayed identity, 0 = the perceived identity is the same as the displayed identity), and then fitted  donation, suspicion, and the displayed identity  with a logistic regression. 
Indeed, we observed a significant interaction effect between the displayed identity and participants' suspicion (p < 0.01). Specifically, people who talked to ``Jessie'' and perceived it as a human were the most likely to donate. When participants talked to ``Jessie (bot)'' but perceiving it as a human, they were also more likely to donate than those in the same condition but perceiving it as a bot. In contrast, when participants talked to ``Jessie'' but suspected it was a bot, they were least likely to make a donation, which supported the UVM in Hypothesis 3, showing people may feel uncomfortable when interacting with a bot with a seemingly human identity. Further, we added the inquiry factor into the model and identified a significant three-way interaction between inquiry, the bot identity and suspicion (p < 0.05). For more details on the interaction effects, please refer to the supplementary material.

Given that the user's suspicion interacted with the experimental conditions, in order to see a clearer result, we decided to reduce the data to the group of participants who believed in the displayed bot identity, and performed the same analysis on \textbf{the reduced dataset}. 
We tested the main effects of bot displayed identity and inquiry on the donation probability using logistic regressions. The result showed that \textbf{``Jessie (bot)'' would decrease the donation probability} ($\beta = -0.52, p<0.05$). So the bot's identity matters in the persuasion outcome, which again disproves  Hypothesis 1.  Although adding the interaction between bot identity and inquiry didn't give significant results,  we did observe some significance using t-tests comparing conditions, shown in Figure~\ref{fig:reduced}. 
In the condition where personal-inquiry was applied, if the ``'Jessie (bot)'' identity was displayed, the donation probability was significantly lower ($p < 0.05$), which indicates that \textbf{when people recognized the bot identity, they didn't like to be asked personal-related questions}. This result is consistent with the UVM in Hypothesis 3 that states there is an interaction effect between the bot identity and persuasive inquiry type on the persuasion outcome.

\subsection{Partner Impression and Conversation Quality}

\begin{figure*}[h]
\centering
\begin{minipage}[b]{.48\textwidth}
  \centering
  \includegraphics[width=\columnwidth]{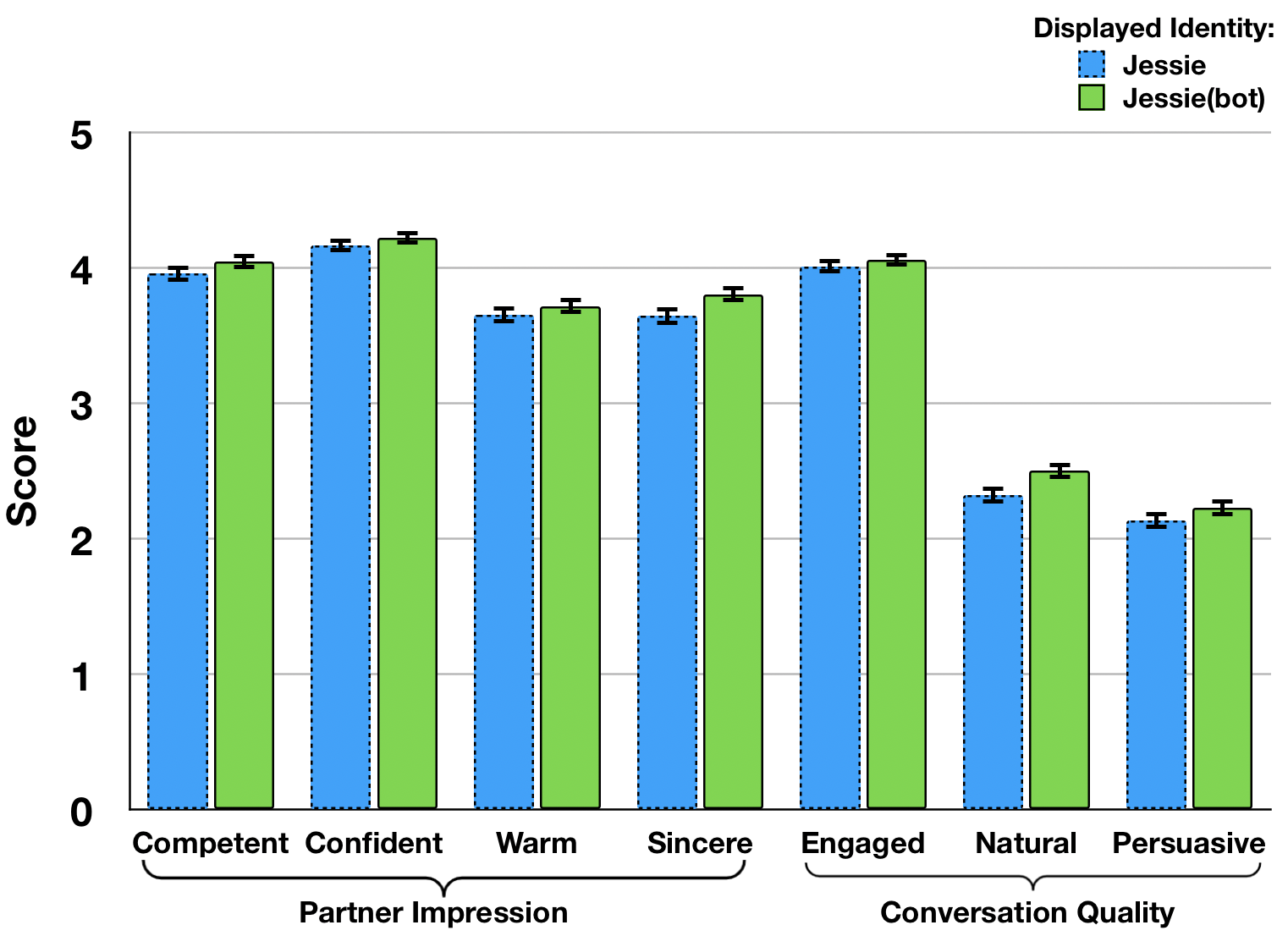}
\caption{Partner impression and conversation quality score under the two \textit{displayed identities} and four persuasive inquiries \textbf{on the whole dataset}.}
\label{fig:natural bot}
\end{minipage}\qquad
\begin{minipage}[b]{.48\textwidth}
  \centering
  \includegraphics[width=\columnwidth]{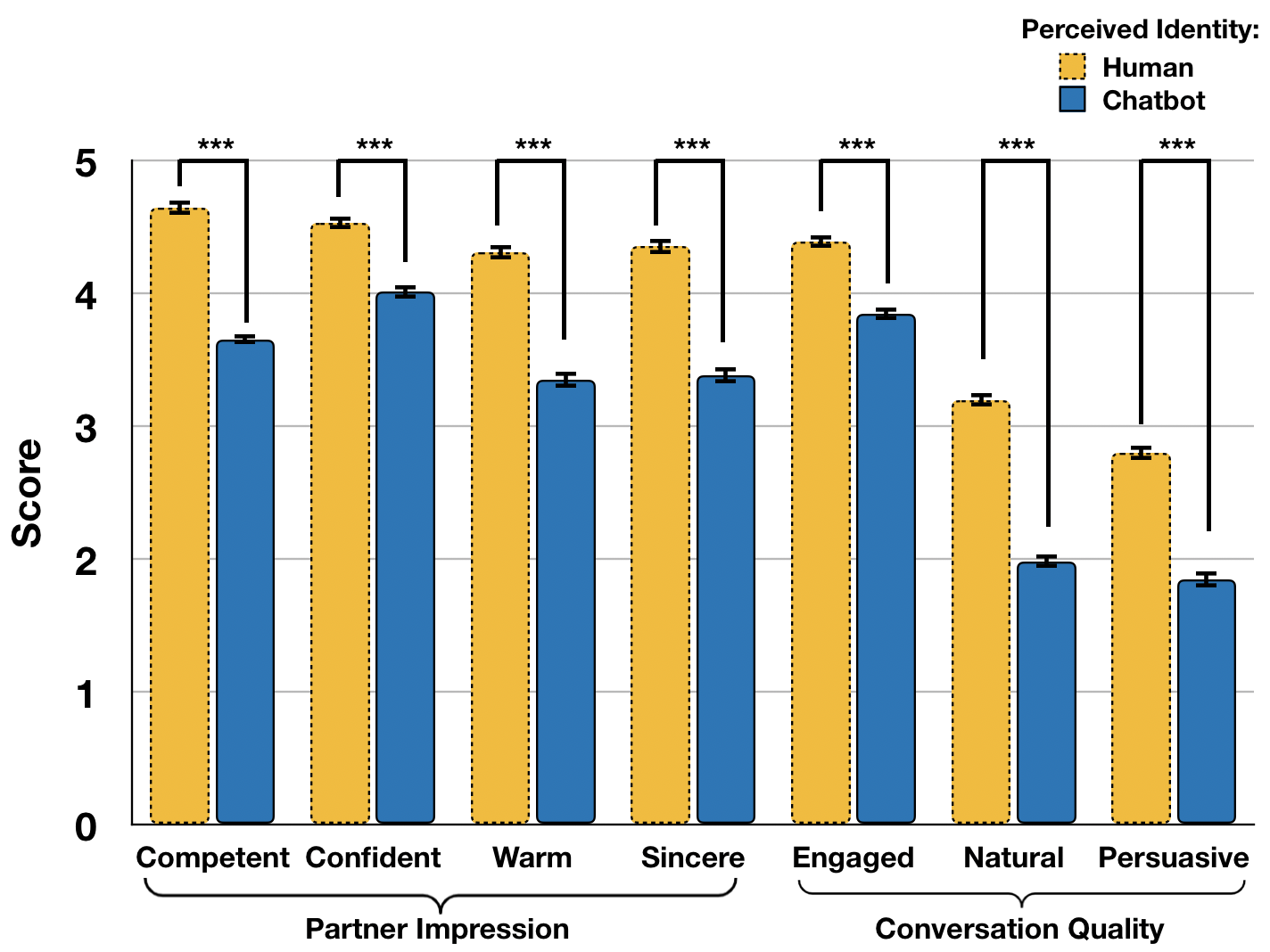}

\caption{Partner impression and conversation quality score under the two \textit{perceived identities} and four persuasive inquiries \textbf{on the whole dataset}.}
\label{fig:natural bot perceived id}
\end{minipage}
\end{figure*}

For dependent variables related to the partner impression (competence, confidence, warmth and sincerity) and conversation quality (engagement, naturalness and persuasiveness), we only observed one significant result: when being presented with the ``Jessie (bot)'' label, participants thought the conversations were more natural ($\beta=0.18, p<0.05$). This suggests participants may have lower expectations of their bot partners when the bot directly discloses its bot identity. Even though the bot conveyed messages with similar quality in two displayed identity conditions, participants still thought the bot's responses were more natural in the disclosed bot identify condition.

Given the previous findings on the impact of bot perception regardless of displayed identities, we were also curious about the perceived identity's impact on the partner impression and conversation quality. We fitted these dependent variables, human perceived identity rather than the displayed identity and the inquiry with a regression model. The results showed that when participants thought they were talking to a human, they would think that the partner was more competent, confident, sincere, and warm, and the conversation was also deemed more engaging, natural, and persuasive ($p<0.001$). 
The average scores of the partner impression and conversation quality are shown in Figure~\ref{fig:natural bot} and \ref{fig:natural bot perceived id} for displayed identity and perceived identity respectively. Figure~\ref{fig:natural bot perceived id} showed many significant results while Figure~\ref{fig:natural bot} didn't, which further confirmed that people's perception and suspicion interacted and meddled the results and removing the suspicion would give clearer outcomes.

Given the interaction effects, we also performed the analysis on the reduced dataset. The regression models showed that the ``Jessie (bot)'' identity had a significantly negative impact on all four impressions and all three conversation quality metrics ($p<0.001$). The findings on partner impression and conversation quality, together with the previous analysis on donation outcome,  showed that using the reduced dataset would give us a clearer picture of the results. Therefore, in the following sections, we discuss all analyses and results based on \textbf{the reduced dataset}.

Since we observed higher donation probability when the perceived identity was human, another natural question to ask is what factors led to a higher probability of perceiving the chatbot as a human regardless of the displayed identity. We used the reduced dataset and fitted the perceived identity (1 = bot, 0 = human) with the conversation quality and impression variables in logistic regression models. The results showed  competent impression, natural responses and persuasiveness all contributed positively to a human identity perception ($p<0.05$). 
This suggest that when building a persuasive bot agent, we need to improve the response and conversation quality.

Further, we were curious to see which interpersonal impressions and conversation quality variables correlated with the donation outcome. Therefore, we performed correlation tests and the results are shown in Table~\ref{tab:cor}. We see that competent and warm impressions had a significantly positive correlation with the donation outcome.

The correlation study led us to wonder, given that the users already perceived the partner as a bot, whether  the competent and warm impressions would still produce positive donation results. We selected the group of people who thought they were talking to a bot, and fitted the donation outcome with the partner impression and conversation quality variables. We found that indeed, even though users already thought they were having conversations with a bot, the more competent the users' impressions were towards the bot partner, the higher the donation probability was ($\beta=0.53, p<0.01$). 

\begin{table}[]
\begin{adjustbox}{width=\columnwidth}
\begin{tabular}{|l|l|l|l|l|l|l|l|l|}
\hline
     & Dona.  & Comp. & Conf. & Warm & Sinc. & Enga. & Natu. & Pers. \\ \hline
Dona.       & 1        & 0.20**       & 0.08     & 0.17*   & 0.12        & 0.03       & 0.13        & 0.15           \\ \hline
Comp.     & --       & 1          & 0.60***       & 0.57***   & 0.57***        & 0.38***       & 0.58***        & 0.51***           \\ \hline
Conf.     & --       & --         & 1          & 0.51***   & 0.54***        & 0.23***       & 0.34***        & 0.35***           \\ \hline
Warm         & --       & --         & --         & 1      & 0.76***        & 0.25***       & 0.48***        & 0.42***           \\ \hline
Sinc.    & --       & --         & --         & --     & 1           & 0.27***       & 0.50***        & 0.43***           \\ \hline
Enga.     & --       & --         & --         & --     & --          & 1          & 0.40***        & 0.36***           \\ \hline
Natu.     & --       & --         & --         & --     & --          & --         & 1           & 0.55***           \\ \hline
Pers.  & --       & --         & --         & --     & --          & --         & --          & 1              \\ \hline
\end{tabular}
\end{adjustbox}
\caption{Correlations between different variables. Donation Probability (Dona.), Competence (Comp.), Confidence (Conf), Warmth; Sincerity (Sinc.), Engagement (Enga.), Naturalness (Natu.), Persuasiveness (Pers.).  *: p < 0.05; **: p < 0.01; ***: p < 0.001. Bonferroni corrections were applied.}\label{tab:cor}
\end{table}
We can conclude from all the previous observations that \textbf{it's not the displayed identity but the perceived identity by the user that matters}; and \textbf{we can improve the system responses and user impression to achieve a greater perception in human identity to facilitate persuasion}; in the cases where we cannot control people's perceived identity, (e.g. fixed as a bot), \textbf{the system's ability to produce a good conversation and impression matters}. So to be effective in persuasion, the chatbot needs to have natural responses, and leave people with good impressions such as being competent.






\subsection{LIWC Linguistics Features}
To explore the participants' linguistics features when facing different perceived identities, we utilized the LIWC (Linguistic Inquiry and Word Count) \cite{liwc} to 
analyze the clue words from different linguistic categories. Positive emotional words occurred more ($p<0.001$) when users thought they were talking to a human. Also, they tended to use more words related to making distinctions \cite{tausczik2010psychological}, such as ``but'' and ``not'', when talking to a chatbot ($p<0.001$). 
These observations suggest that users may be more likely to express feelings when they perceived the partner as  a human, and feel the need to make distinctions more when they perceived the partner as a chatbot.




\subsection{Inconsistent Donation Behavior}
In our task, participants were asked to input their donation amount privately after the conversation, and this amount would be their actual donation to the charity deducted from their task payment. However, similar to the findings in \cite{wang2019persuasion}, we noticed that people didn't always keep their donation promise. For example, they would indicate in the conversation that they wanted to donate \$2 but donated \$0 in the end. We want to investigate if the inconsistent donation behavior is related to the identity disclosure or the persuasive inquiry. Therefore, we defined a new dependent binary variable to indicate the inconsistent donation behavior (1 = the actual donation amount is not equal to the amount indicated in the conversation, 0 = the two amounts are the same). We fitted this variable along with the experimental conditions into a logistic regression model on the reduced dataset.

The analysis showed that {\bf if participants thought  they were talking to a human, they were more likely to have inconsistent donations ($\beta=0.70, p<0.01$)}. This seemingly surprising result may be explained by different interaction dynamics. One possible explanation is that when talking to other humans, people may have experienced the innate social pressure to save face and act generously by indicating a bigger donation, while such pressure seem to disappear in the context of human-bot interactions. We also observed that if the perceived identity was human, personal inquiry interacted with the human ``Jessie'' identity and led to more consistent behaviors ($\beta=-1.52, p<0.05$), compared to the ``no inquiries'' condition. Such finding may also be explained by interaction dynamics, such that personal inquiry shortened the interpersonal distance between the two parties, encouraging the participants to be more consistent.

We also found that participants' personality impacted the inconsistent donation behavior: participants who were more rational in the Decision-Making style test and endorsed purity more in the Moral Foundation endorsement test were more likely to keep their promises ($p<0.05$).

\subsection{Donation Behavior and Personality}
People have concerns about the deployment and misuse of such persuasive technologies in the real-world, since persuasion is to influence people's thoughts and change their behaviors. Therefore, we are curious to see what kind of people may be more easily persuaded than others in order to prepare for future countermeasures of unethical persuasion. We fitted the user profile information and the donation outcome with a logistic regression model on the reduced dataset. 

The analysis showed being male ($\beta = 0.62, p < 0.05$), having medium income ($\beta = 0.81, p < 0.01$) and having the ``caring'' ($\beta = 0.42, p < 0.05$) trait were significant predictors of donation. We further controlled these variables in the models of the main analyses on donation outcome, bot identity and inquiry, and found the results are consistent with previous findings.







\section{Ethical Considerations}
Persuasion is a double-edged sword and has been used for good or evil throughout the history. Developing persuasive chatbots is a very new research field and needs careful thinking in design and deployment ethics. As the nation's first autobot regulation, California's new Autobot Law~\cite{calautobotlaw} (effective July 1, 2019) enforces the identity disclosure of artificial chatbot in commercial uses. Our findings provide important insights in light of the bot regulation. The finding that some people still perceive an identified bot as a human is concerning. As the technology advances and human-bot conversations become more pervasive and natural, people may become less discerning regarding bots' identities, even when they are informed. In another aspect, our findings also suggest disclosing bot's identity is necessary because when people do see the bot identity, they are more reactant to personal inquiries and evaluate conversations that are more personal as less persuasive. This reactance is protective when considering potential abusive uses of persuasive chatbots. Given these, the chatbot research community needs to conduct more studies to further test these phenomena. 

From a bigger perspective, regardless of the bot's communication capacity or identity disclosure, we suggest when designing a persuasive chatbot, the designers ought to first define the bot intention in influencing human perceptions and behaviors. Invoking Quintilian's rhetoric, we ought to build ``a good chatbot speaking well." The `good" and `well" mean the bot is designed to benefit the users and the society at large through democratic communicative processes, such that users are ensured with information accuracy, transparency, and autonomy in making up their own mind and decisions. In this regard, our current persuasive chatbot is designed with a clear well-intended persuasive goal, that can be integrated with other forms of persuasive campaigns. However, there are other contexts that require more cautions when introducing persuasive chatbots as they involve more complex or contentious intentions, such as changing certain health (e.g., vaccination) or political (e.g., voting) behaviors. Discussing their ethical boundaries is beyond the scope of the current paper, but we want to stress that the design of chatbot ought to follow ethical guidelines established in the persuasion literature and preemptively consider human's vulnerability in communication, such as the identity confusion identified in our research.  

\section{Limitations and future work}
One primary limitation is the bot's conversation quality. The chatbot follows a rigid agenda to stay on task and uses template-based or retrieval-based responses, so it's  not flexible in responding to undefined scenarios and occasionally conveys incoherent responses. 
The chatbot's limited quality is also reflected in the reasons of confusion shown in Table~\ref{tab:reason}. 
We plan to explore neural generation methods to improve conversation quality and encourage other researchers to explore our data. 
A second limitation concerns the identity manipulation. Our current manipulations were conservative because we only used a simple cue (``Jessie (bot)'') to elicit the bot identity. Future research could experiment with a stronger manipulation and we anticipate the effects identified here would be strengthened. Another limitation concerns the conversation's length. 
 To be efficient in the data collection, 
 participants could choose to continue or exit the conversation after 10 conversational turns or if the system promptly closed the conversation.
Future research should consider testing longer conversations to explore more complex conversation dynamics. Finally we are aware of the limitation in the generalizability of the findings identified in a specific online donation context. Future research should test the hypotheses in other persuasion domains and we expect the core theoretical propositions can be replicated across domains.

\section{Conclusions}
Intelligent chatbots have been increasingly adopted in various applications, using different identities. 
Our study investigated how the bot's identity and persuasive inquiries influence persuasion outcome. We recruited 790 participants to interact with our agenda-based chatbot on a donation persuasion task and designed a two by four factorial experiment with hidden or disclosed bot identity and four inquiry strategies to test our hypotheses. We found that it's not the displayed identity  but the  identity perceived  by people that impacted the donation outcome. People would suspect the identity displayed to them and such suspicion influenced donation outcomes. After removing participants who suspected the displayed identity, we did see that disclosing the bot's identity decreased the donation probability. Also personal inquiry worked better on people who thought they were talking to a human. Another take-away is that the chatbot's quality highly correlated with donation outcomes. We should strive to achieve a more human-like chatbot by improving the response naturalness. 
Our published code, data and analyses serve as the first step to build effective persuasive dialogue systems.

\section{Acknowledgments}
This work was supported by an Intel research gift. We thank all the reviewers for their helpful feedbacks.

\bibliographystyle{SIGCHI-Reference-Format}
\balance
\bibliography{sample}


\appendix

\begin{table*}[ht]
\small
\centering
\begin{adjustbox}{width=0.95\textwidth}
\begin{tabular}{l|l}
\hline

\textbf{Persuasion Appeal}   & \textbf{Example} \\\hline
Logical appeal                 & \textit{\begin{tabular}[c]{@{}l@{}}Your donation could possible go to this problem and help many young children. \\ You should feel proud of the decision you have made today.\end{tabular}}                                     \\\hline
Emotion appeal               & \textit{\begin{tabular}[c]{@{}l@{}}Millions of children in Syria grow up facing the daily threat of violence.  \\ This should make you mad and want to help.\end{tabular}}            \\\hline

Credibility appeal             & \textit{\begin{tabular}[c]{@{}l@{}}And the charity is highly rated with many positive rewards.\\ You can find reports associated with the financial information by visiting this link.\end{tabular}}                             \\\hline

Foot-in-the-door      & \textit{\begin{tabular}[c]{@{}l@{}}And sometimes even a small help is a lot, thinking many others will do the same.\\ By people like you, making a a donation of just \$1 a day, you can feed a child for a month.\end{tabular}} \\\hline

Self-modeling     & \textit{\begin{tabular}[c]{@{}l@{}}I will  donate to Save the Children myself. \\ I will match your donation.\end{tabular}}                                                                                                      \\\hline
Personal story                 & \textit{\begin{tabular}[c]{@{}l@{}}I like to give a little money to charity each month. \\ My brother and I replaced birthday gifts with charity donations a few years ago.\end{tabular}}                                        \\\hline
Donation information & \textit{\begin{tabular}[c]{@{}l@{}}Your donation will be directly deducted from your task payment. \\ The research team will collect all donations and send it to Save the Children.\end{tabular}} \\\hline

\end{tabular}
\end{adjustbox}
\caption{Example sentences for the seven persuasive appeals from \protect\cite{wang2019persuasion} 
}
\label{tab:sample sentence for strategy}
\end{table*}

\begin{figure*}
\centering
  \includegraphics[width=1.3\columnwidth]{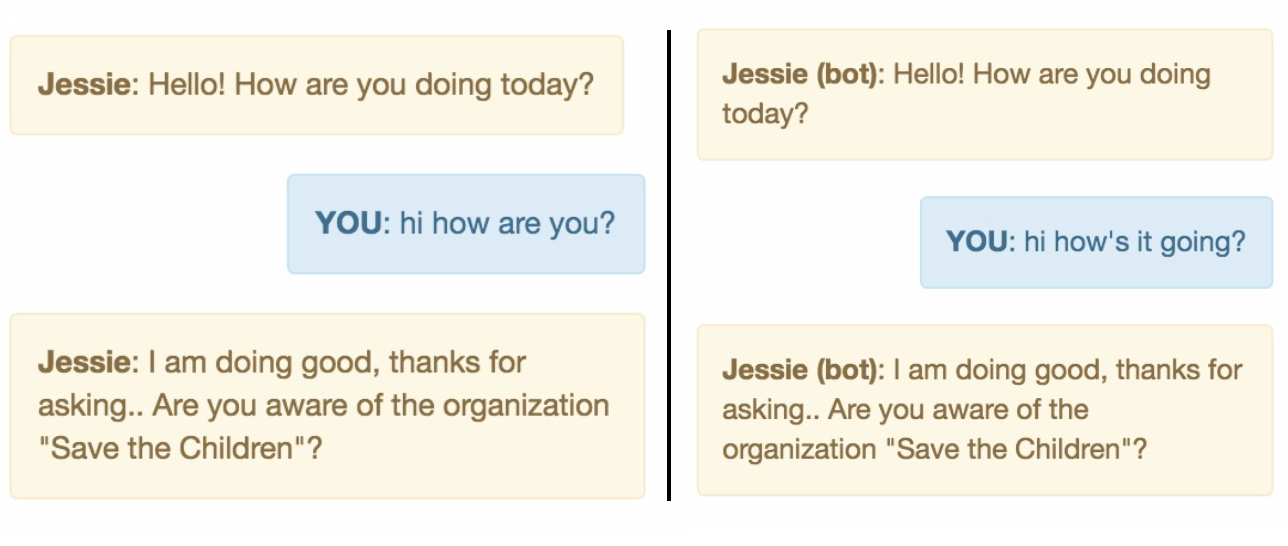}
  \caption{We have ``Jessie" and ``Jessie (bot)'' in front of the system utterances to indicate chatbot's identities in different experimental conditions.}
  
 \label{fig:identities}
\end{figure*}


\begin{figure*}
    \centering
    \includegraphics[width=1.8\columnwidth]{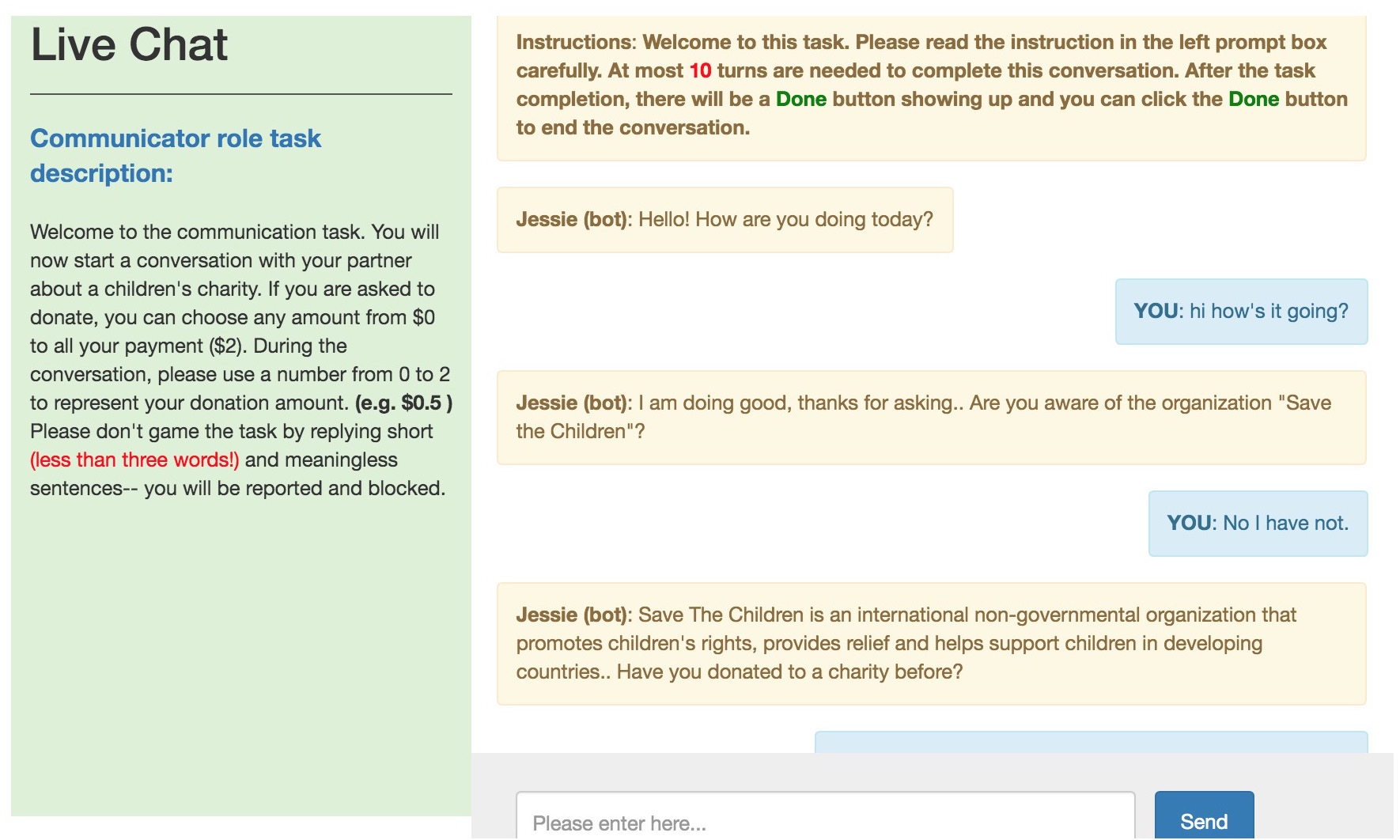}
    \caption{Chat interface.}
    \label{fig:my_label}
\end{figure*}

\begin{figure*}
\centering
\begin{minipage}[b]{.48\textwidth}
  \centering
  \includegraphics[width=\columnwidth]{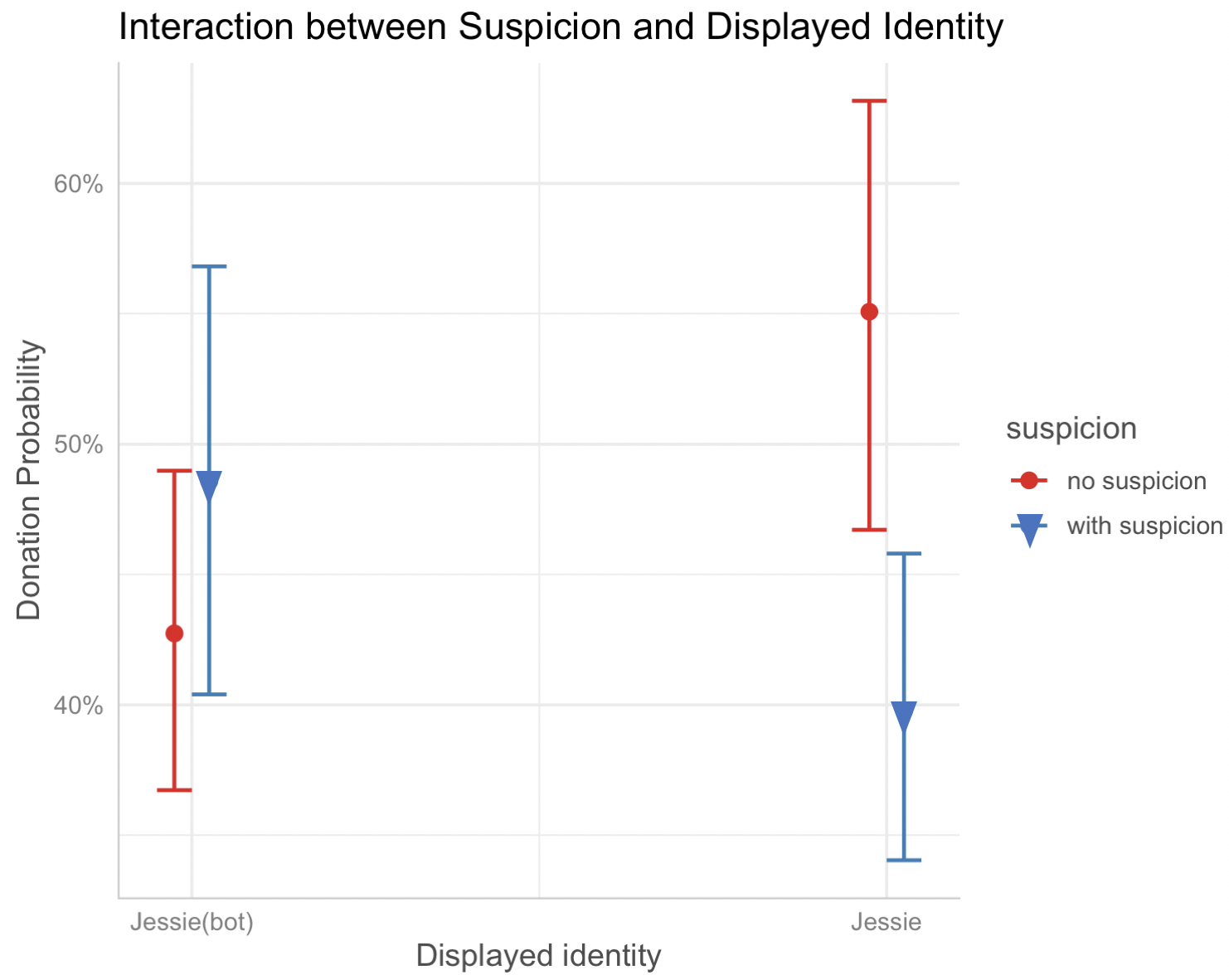}
\caption{Interaction effects between  suspicion and the displayed identity.}
\label{fig:inter2}
\end{minipage}\qquad
\begin{minipage}[b]{.48\textwidth}
  \centering
  \includegraphics[width=\columnwidth, height=6.8cm]{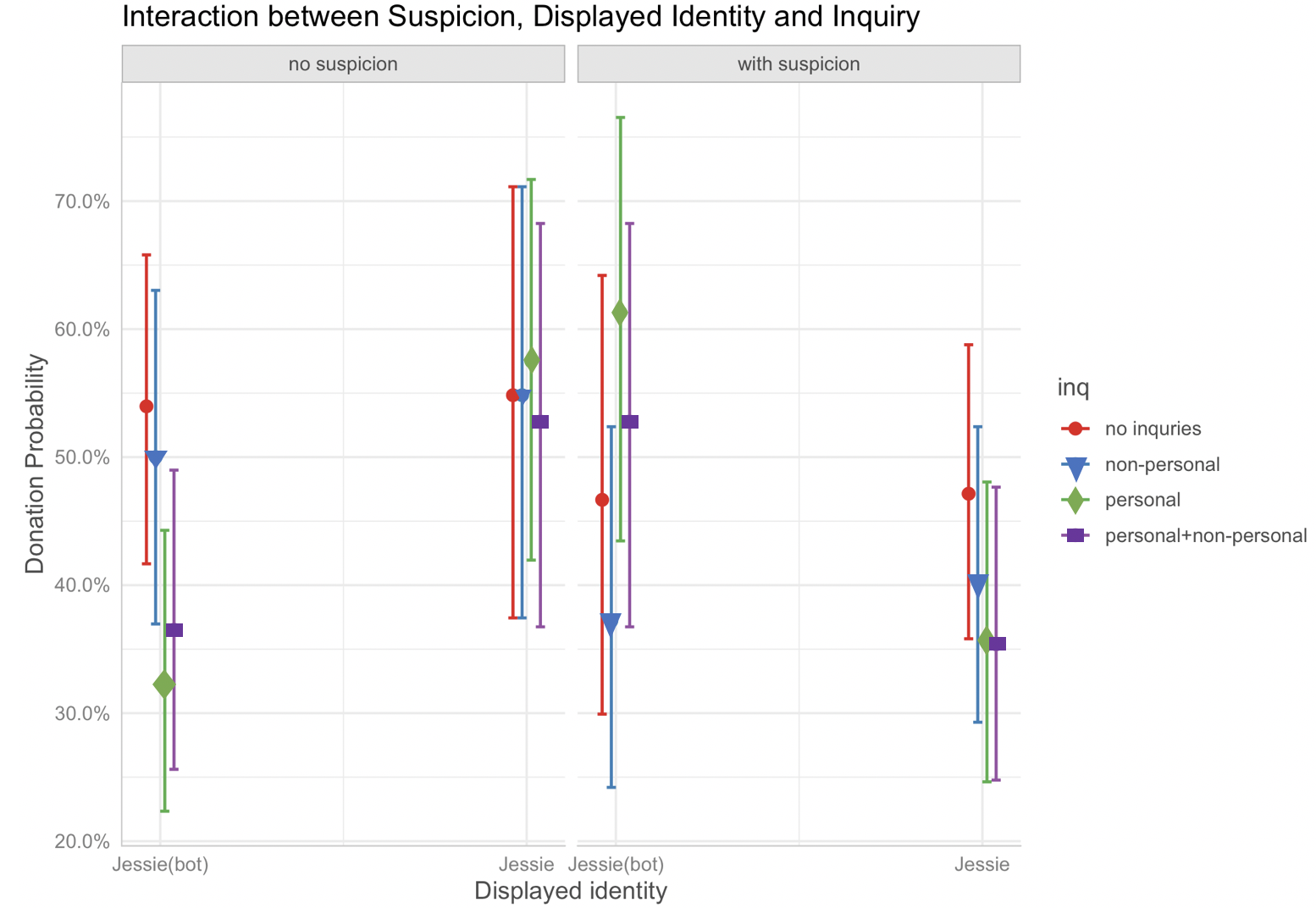}

\caption{Three-way Interaction effects between suspicion, the displayed identity, and the persuasive inquiry.}
\label{fig:inter3}
\end{minipage}
\end{figure*}

\end{document}